\documentclass[aip,jcp,reprint,superscriptaddress]{revtex4-1}
\usepackage{amsfonts}
\usepackage{amsmath,amssymb,epsfig,color}
\usepackage{float}
\usepackage{graphicx}
\usepackage{graphicx,color}
\usepackage{setspace}
\usepackage{hyperref}
\usepackage{textgreek}
\usepackage{soul}
\usepackage{subfig}
\usepackage{natbib}

\newcommand{\shortrightarrow}[1][3pt]{\mathrel{%
   \hbox{\rule[\dimexpr\fontdimen22\textfont2-.2pt\relax]{#1}{.4pt}}%
   \mkern-4mu\hbox{\usefont{U}{lasy}{m}{n}\symbol{41}}}}

\begin{document}

\title{Assessing the source of error in the Thomas-Fermi-von Weizs\"acker density functional}
\author{Bishal Thapa}
\affiliation{Department of Physics and Astronomy, George Mason University, Fairfax, VA 22030, USA}
\affiliation{Quantum Science and Engineering Center, George Mason University, Fairfax, VA 22030, USA}
\author{Xin Jing}
\affiliation{College of Engineering, Georgia Institute of Technology, Atlanta, GA 30332, USA}
\affiliation{College of Computing, Georgia Institute of Technology, Atlanta, GA 30332, USA}
\author{John E. Pask}
\affiliation{Physics Division, Lawrence Livermore National Laboratory, Livermore, California 94550, USA}
\author{Phanish Suryanarayana}
\affiliation{College of Engineering, Georgia Institute of Technology, Atlanta, GA 30332, USA}
\author{Igor I. Mazin}
\email[Email: ]{imazin2@gmu.edu}
\affiliation{Department of Physics and Astronomy, George Mason University, Fairfax, VA 22030, USA}
\affiliation{Quantum Science and Engineering Center, George Mason University, Fairfax, VA 22030, USA}
\date{\today }

\begin{abstract}
We investigate the source of error in the Thomas-Fermi-von Weizs\"acker (TFW) density functional relative to  Kohn-Sham density functional theory (DFT). In particular, through numerical studies on a range of materials, for a variety of crystal structures subject to strain and atomic displacements, we find that while the ground state electron density in TFW orbital-free DFT is 
close to the Kohn-Sham density, 
the corresponding energy deviates significantly from the Kohn-Sham value. We show that these differences are a consequence of the poor representation of the linear response within the TFW approximation for the electronic kinetic energy, confirming conjectures in the literature. In so doing, we find that the energy computed from a non-self-consistent Kohn-Sham calculation using the TFW electronic ground state density is in very good agreement with that obtained from the fully self-consistent Kohn-Sham solution. 
\end{abstract}

\maketitle

\section{Introduction} \label{Sec:Introduction}
Density functional theory (DFT) \cite{martin2020electronic,  parr1980density} is one of the most widely used ab initio methods in physical, chemical, and materials science research for understanding and predicting the properties of materials systems.  Its conceptual foundation lies in the Hohenberg-Kohn (HK) theorem\cite{PhysRev.136.B864}, which states that the total energy of the system is a unique, albeit unknown, functional of its density.   This rather formal mathematical concept was made practically useful  by the Kohn-Sham (KS) formalism \cite{kohn1965self},  wherein the real  system of interacting electrons is replaced by a fictitious system of non-interacting fermions that generates the same electronic density.  In particular, the electronic kinetic energy is no longer an explicit functional of the density, but rather takes the form (in atomic units):
$$
T_s =  - \frac{1}{2} \sum_{n=1}^{N_s} \int \psi_n (\mathbf{r}) \nabla^2 \psi_n(\mathbf{r}) \mathrm{d \mathbf{r}} \,,
$$
where $\{\psi_n \}_{n=1}^{N_s}$ are the KS orbitals.   In so doing, a one-electron Schr{\"o}dinger-type equation needs to be solved for multiple electronic states,  i.e., KS orbitals, whose number grows with the system size, which together with the orthogonality constraint on the orbitals, results in computations that scale cubically with system size \cite{martin2020electronic}.  This severely restricts the range of systems that can be studied using KS-DFT. 
 
An alternative to replacing the system of interacting electrons with a fictitious system of non-interacting fermions is to replace it instead with a fictitious system of non-interacting bosons. This can be achieved by approximating the kinetic energy $T_s$ using an explicit functional of the density, the resulting formalism referred to as orbital-free (OF) DFT \cite{ligneres2005introduction}.  This  generally amounts to solving a Schr{\"o}dinger-type equation for only one electronic state, which corresponds to the square root of the density. Though the computational cost of OF-DFT scales linearly with system size, thus overcoming the cubic-scaling bottleneck of KS-DFT, it is rarely used in practice due to the lack of accurate approximations for $T_s$. 
 
Historically, the first approximation for $T_s$ was suggested long before the advent of DFT, by Thomas \cite{thomas1927calculation} and  Fermi \cite{fermi1928statistische}:  
 $$
 T_{TF}=\int \left( \frac{3}{10}(3\pi ^{2})^{\frac{2}{3}}\rho({\mathbf r})^{\frac{5}{3}} \right) \mathrm{d{\mathbf r}}, 
 $$
which is exact is the limit of slowly varying density. The first gradient correction: 
$$
T_{\lambda W}=\int \left(\frac{\lambda}{8} \frac{\vert \nabla \rho(\mathbf{r})\vert^2}{\rho(\mathbf{r})} \right) \mathrm{d}{\mathbf r},
$$ 
with the weight factor of $\lambda = 1$ was derived by von Weizs\"acker \cite{werzsticxer1935theorie}. This term by itself represents a lower bound on $T_s$, being  exact in the limit of small and rapid (large wavevector) density variations \cite{jones1971density}. A similar formula with weight factor of $\lambda= 1/9$  was derived by Kirzhnits \cite{kirzhnits1957quantum}, the kinetic energy $T_s = T_{TF} + T_{\lambda W}$ being  exact in the opposite limit of slow but not necessarily small variations. Realizing the importance of satisfying the constraint arising from the aforementioned lower bound, a number of other local kinetic energy functionals have recently been proposed, including those of the generalized gradient approximation (GGA) \cite{luo2018simple, constantin2018semilocal, francisco2021analysis} and the Laplacian-level meta-GGA \cite{perdew2007laplacian, constantin2017modified} varieties. Though these functionals have found significant success, they are semi-empirical in that they involve parameters that could be material dependent. There have also been suggestions to  use alternate values for the weight factor \cite{parr1980density}, which can be made position dependent  \cite{tomishima1966solution}, however such strategies also depend upon the class of system under consideration and so lack universality. 

The dramatic difference between the limiting values for the weight factor in $T_{\lambda W}$ signals the inability of the local TFW functional, i.e.,  $T_s = T_{TF} + T_{\lambda W}$, to describe the kinetic energy  even for small density variations, if the scale of variations is neither particularly large nor particularly small. This led  to thinking, as early as three decades ago \cite{chacon1985nonlocal},  of the need for functionals that are nonlocal in coordinate space, i.e., depend on the density correlation at finite distances, e.g.:
$$
T_s =T_{TF}+\int \rho(\mathbf{r})K(\rho(\mathbf{r}),\rho(\mathbf{r}')) \rho(\mathbf{r}') \, \mathrm{d \mathbf{r}} \, \mathrm{d\mathbf{r}'} \,,
$$
where the nonlocal kernel $K(\rho(\mathbf{r}),\rho(\mathbf{r}'))$, is selected in such a way so as to
improve the description of the noninteracting susceptibility
$\chi_{0}^{-1}(\mathbf{r,r}^{\prime})={\delta^{2}T_{{s}}}/{\delta\rho
(\mathbf{r)}\delta\rho(\mathbf{r}^{\prime}\mathbf{)}}$ to reproduce the uniform electron gas limit given by the analytic Lindhard formula \cite{ziman1972principles}.
%: \rev{$\chi_{Lind}^{-1}(\mathbf{r,r}^{\prime})=-{\delta^{2}T_{{s}}}/{\delta\rho(\mathbf{r)}\delta\rho(\mathbf{r}^{\prime}\mathbf{)}}$}. 
Note that a slightly different form, essentially equivalent to that above, has also been proposed \cite{chacon1985nonlocal}: $T_s=T_{TF}+T_{W}[\tilde\rho(\mathbf{r})]$, where
$\tilde\rho(\mathbf{r})=\int \rho(\mathbf{r}')K(\rho(\mathbf{r}),\rho(\mathbf{r}'))\, \mathrm{d\mathbf{r}'}$. In subsequent work \cite{Igor}, it was pointed out that a form more consistent with the idea of generalizing the von Weizs\"acker-Kirzhnits term into the domain of nonlocal functionals should include density log-gradients in powers adding up to 2, e.g., 
$$
T_s = T_{TF} + \int \frac{\nabla \rho(\mathbf{r}) }{\rho(\mathbf{r})}K(\rho(\mathbf{r}),\rho(\mathbf{r}')) \frac{\nabla \rho(\mathbf{r}') }{\rho(\mathbf{r}')}  \mathrm{d \mathbf{r}} \, \mathrm{d\mathbf{r}'} \,.
$$
A similar idea was later proposed by Wang and Teter \cite{wang1992kinetic}, who represented the kinetic energy  as
$$
T_s = T_{TF}+ T_{\lambda W}+\int {\rho(\mathbf{r})^aK(\mathbf{r},\mathbf{r}') }{\rho(\mathbf{r}')}^b \mathrm{d \mathbf{r}} \, \mathrm{d\mathbf{r}'},
$$
where $a+b=5/3$. This was later generalized onto density-dependent kernels by Wang, Govind, and Carter (WGC) \cite{wang1999orbital}:
$$
T_s = T_{TF}+ T_{\lambda W}+\int {\rho(\mathbf{r})^aK(\rho(\mathbf{r}),\rho(\mathbf{r}')) }{\rho(\mathbf{r}')}^b \mathrm{d \mathbf{r}} \, \mathrm{d\mathbf{r}'} \,,
$$
where $a+b = 8/3$. 
While showing good results for particular problems \cite{wang1999orbital, carling2003orbital, zhou2005improving, ho2007energetics, huang2010nonlocal,  shao2021revised, doi:10.1063/1.5023926, xu2022nonlocal}, such nonlocal functionals have found rather limited use in practice due to greater computational expense and the need for specialized kernels to be developed for different materials systems \cite{zhou2005improving, huang2010nonlocal, doi:10.1063/1.5023926, shin2014enhanced, shao2021revised}. Further advances require a more fundamental understanding of linear response, and in particular, whether it is the deciding factor in determining the error associated with OF-DFT, as conjectured in previous works \cite{chacon1985nonlocal, wang1992kinetic,Igor, mazin1998nonlocal}. 

An alternative strategy to nonlocal density functionals is to perform a non-self-consistent KS calculation with the electronic ground state density from OF-DFT as input. In particular, the ground state electron density obtained from an OF-DFT calculation is used as input to  a single self-consistent field (SCF) iteration of a KS calculation, the quantities so obtained then being used to compute the energy, either using the Kohn-Sham functional \cite{ullmo2001semiclassical}, the well-studied \cite{robertson1991does, farid1993extremal}  Harris-Foulkes functional \cite{harris1985simplified, foulkes1989tight} as in the shell correction method \cite{yannouleas1998energetics}, or a weighted combination of the two as in the orbital-corrected orbital-free (OO) DFT \cite{zhou2006orbital, zhou2008accelerating}. The formulations/implementations \cite{yannouleas1998energetics, ullmo2001semiclassical, zhou2008accelerating}  related to the Strutinsky shell correction method \cite{strutinsky1968shells} are for molecular systems in the context of TF OF-DFT,  while the focus in the current work is on TFW and condensed matter systems. The OO-DFT method, developed in the context of WGC OF-DFT, displays very good agreement with KS-DFT.  Though interesting, it was, however, semi-empirical, with dependence on a parameter that determines the relative contributions of the Kohn-Sham and Harris-Foulkes energies, in contrast to the above-mentioned nonlocal functionals underpinned by linear response theory. Furthermore,  WGC already incorporates some amount of linear response, complicating the ability to draw clear inferences.  Finally, the numerical evidence consists of just two cases, namely the energy-volume curves for fcc Ag and cubic-diamond Si. To what extent this idea is generally applicable, and how (if at all) it is related to linear response theory, remain to be clarified.
%, which provides motivation for the present work. 

In this paper, we address the aforementioned issues. First, we verify that the ``single-shot'' KS calculation with TFW OF-DFT ground state density as input is close to the fully self-consistent KS-DFT solution for a range of materials, including different crystal structures subject to volumetric  and symmetry-lowering perturbations. Second, we argue that the success of such a strategy indicates that the main limitation of TFW OF-DFT is indeed its inability to properly describe the correct linear response, as conjectured in the literature \cite{chacon1985nonlocal, wang1992kinetic,Igor, mazin1998nonlocal}.  
 
%%%%%%%%%%%%%%%%%%%%%%%%%%%%%%%%%%%%%%%%%%%%%%%%%%%%
%%%%%%%%%%%%%%%%%%%%%%%%%%%%%%%%%%%%%%%%%%%%%%%%%%%%
\section{Systems and Methods}
We consider body-centered cubic (BCC), face-centered cubic (FCC), hexagonal close packed (HCP),  and body-centered tetragonal (BCT) crystals of magnesium (Mg), aluminum (Al),  and indium (In); as well as diamond cubic (DC) and hexagonal diamond (DH) (also known as lonsdaleite) crystals of silicon (Si). These systems form a diverse set that includes a simple metal, transition metal, and semiconductor, in a variety of lattice configurations.  Importantly, well-tested local pseudopotentials \cite{zhou2004transferable} are available for the chemical elements in question, i.e., Mg, Al, In, and Si, allowing for a careful comparison of the results obtained from KS-DFT and OF-DFT calculations. 

Unless specified otherwise, we choose  the primitive unit cells for each of the systems, i.e.,  1-atom cells for the BCC and FCC lattices,  2-atom cells for the HCP, BCT, and DC lattices, and 4-atom cells for the DH lattice.  For the HCP and BCT lattices, we choose the ideal $c/a$ ratios $1.633$ and $1.414$, respectively. After determining the equilibrium configurations,  we consider the following strains and atomic displacements: \vspace{-2mm}
\begin{itemize}
\item Volumetric strains for each of  the aforementioned systems. In this case, the strain tensor takes the form:
\begin{equation}
\mathbf{G} = 
  \begin{pmatrix}
    1+g & 0 & 0 \\
    0 & 1+g & 0 \\
    0 & 0 & 1+g 
  \end{pmatrix} \,,
\end{equation}
where the perturbation parameter $-0.3\leq g \leq 0.3$, with $g < 0$ and $g>0$ corresponding to the contraction and expansion of the unit cell, respectively.
\item \vspace{-2mm} Symmetry-lowering, volume-preserving rhombohedral strains for Mg, Al, and In in the FCC and BCC crystal  configurations; and Si in the DC crystal configuration. In this case, the strain tensor takes the form: 
\begin{equation}
\mathbf{G} = 
  (1+ 3 g)^{-\frac{1}{3}}
  \begin{pmatrix}
    1+g & g & g\\
    g & 1+g & g\\
    g & g & 1+g
  \end{pmatrix} \,,
\end{equation}
where the perturbation parameter $-0.1 \leq g \leq 0.1$, with $g < 0$ and $g>0$ corresponding to the compression and elongation of the unit cell, respectively, along the $\left[ 1 1 1\right] $ direction. 
\item \vspace{-2mm} Volume-preserving uniaxial strains along the $\left[ 0 0 1\right] $ direction for Mg, Al, and In in the HCP and BCT crystal  configurations; and Si in the DH crystal configuration. In this case, the strain tensor takes the form:
\begin{equation}
\mathbf{G} = \begin{pmatrix}
    (1+g)^{-\frac{1}{2}} & 0 & 0\\
    0 &  (1+g)^{-\frac{1}{2}} & 0\\
    0 & 0 & 1+g
  \end{pmatrix} \,,
\end{equation}
where the perturbation parameter $-0.1 \leq g \leq 0.1$, with $g < 0$ and $g>0$ corresponding to the compression and elongation of the unit cell, respectively.
\item \vspace{-2mm} Symmetry-lowering atomic perturbations (i.e, frozen phonons) for Al and Mg in the BCC, FCC, and HCP crystal configurations; and In in the FCC, HCP, and BCT crystal configurations. For BCC and FCC, we choose the 2-atom conventional cell and 2-atom tetragonal cells, respectively, rather than the 1-atom primitive cell used in other simulations. In all cases, the atom is perturbed along the $\left[ 0 0 1\right] $ direction, with the  $z$ coordinate of one of the atoms changed as: $z \rightarrow (1+g) z$, where the perturbation parameter $-0.05 \leq g \leq 0.05$. \vspace{-2mm}
\end{itemize}
The systems considered here ensure that both slow and rapid variations in the density are encountered, especially when the internal parameters are varied, enabling a thorough analysis  of the error in the TFW functional. Note that even in the case of uniform strains, there are rapid changes in the electron density for many of the systems, as evidenced by the large errors in the TFW OF-DFT energies (Supplementary Material).

All calculations are performed using the  M-SPARC code \cite{xu2020m, zhang2023version}, which is a \texttt{Matlab} version of the large-scale parallel electronic structure code, SPARC \cite{xu2021sparc}.  It employs the real-space finite-difference method, whose formulation and implementation in the context of KS-DFT and OF-DFT can be found in previous works \cite{ghosh2017sparc2, ghosh2017sparc1, ghosh2016higher, suryanarayana2014augmented}. We employ the local density approximation (LDA) \cite{kohn1965self,  perdew1981self} for the exchange-correlation functional and use the bulk-derived local pseudopotentials (BLPS) \cite{zhou2004transferable}.  In the OF-DFT calculations, we choose the TFW kinetic energy functional with weight factor $\lambda=1/8$\footnote{We have also tested other choices of $\lambda$, and have found that the final conclusions remain unchanged.}.  In the KS-DFT calculations, we perform Brillouin zone integration using a $15\times15\times15$ Monkhorst-Pack grid for the FCC, BCC, DC, and DH lattices,  and $15\times15\times10$ grid for the HCP and BCT lattices, which ensures that the energies are converged to within $10^{-4}$ ha/atom.  In all calculations, we employ a 12-th order finite-difference approximation and a grid spacing of 0.4 bohr, which ensures that the computed energies are converged to within $10^{-3}$ ha/atom. Finally, the change in energy  arising due to a perturbation, which is the main quantity of interest in the present work (Section~\ref{Sec:Results}), is converged to within $10^{-6}$ ha/atom. 

%%%%%%%%%%%%%%%%%%%%%%%%%%%%%%%%%%%%%%%%%%%%%%%%%%%%
%%%%%%%%%%%%%%%%%%%%%%%%%%%%%%%%%%%%%%%%%%%%%%%%%%%%

\section{Results and discussion} \label{Sec:Results}
We use the framework described in the previous section to perform KS-DFT and OF-DFT calculations for the selected systems. In particular, for each system, we compute the four energies listed below.
\begin{itemize}
\item \vspace{-1.5mm} $E_{KS \shortrightarrow KS} := E_{KS}(\rho_{KS})$: KS-DFT energy $E_{KS}$ corresponding to the KS-DFT ground state density $\rho_{KS}$. This is obtained by performing a standard electronic ground state calculation in KS-DFT.
\item \vspace{-2mm} $E_{OF \shortrightarrow KS} := E_{KS}(\rho_{OF})$: KS-DFT energy $E_{KS}$ corresponding to the OF-DFT ground state density $\rho_{OF}$. This involves the calculation of orbitals for the given $\rho_{OF}$, i.e.,  a single self-consistent field (SCF) iteration in KS-DFT. 
\item \vspace{-2mm} $E_{KS \shortrightarrow OF} := E_{OF}(\rho_{KS})$: OF-DFT energy $E_{OF}$ corresponding to the KS-DFT ground state density $\rho_{KS}$.
\item \vspace{-2mm} $E_{OF \shortrightarrow OF} := E_{OF}(\rho_{OF})$: OF-DFT energy $E_{OF}$ corresponding to the OF-DFT ground state density $\rho_{OF}$. This is obtained by performing a standard electronic ground state calculation in OF-DFT. \vspace{-1mm}
\end{itemize}

To quantify the error in the energies $E_{OF \shortrightarrow KS}$, $E_{KS \shortrightarrow OF}$, and $E_{OF \shortrightarrow OF}$, the error being defined with respect to $E_{KS \shortrightarrow KS}$, we define the following root-mean-square measure, referred to as the $\Delta$-value \cite{lejaeghere2016reproducibility}: 
\begin{equation}
\Delta  = \sqrt{\dfrac{ \int (\Delta E(g) - \Delta E_{KS \shortrightarrow KS}(g))^2 \, dg}{\int dg}} \,,
\end{equation}
where 
\begin{align}
\Delta E(g) & = E(g) - E(0) \,, \nonumber \\
\Delta E_{KS \shortrightarrow KS}(g) & = E_{KS \shortrightarrow KS}(g) - E_{KS \shortrightarrow KS}(0)  \,, \nonumber 
\end{align}
with $E \in \{E_{OF \shortrightarrow KS}, E_{KS \shortrightarrow OF}, E_{OF \shortrightarrow OF} \}$, and the corresponding $\Delta \in \{\Delta_{OF \shortrightarrow KS}, \Delta_{KS \shortrightarrow OF}, \Delta_{OF \shortrightarrow OF} \}$. Note that the difference in energies from $g=0$  is used in the definition of the error since the reference energy within KS-DFT and OF-DFT is different, a consequence of the different energy functionals, i.e., only differences in energy within the same level of theory are meaningful. We emphasize that we intentionally gauge the OF functional against the KS functional, and not against the experiment.

\begin{figure*}[t]
\includegraphics[keepaspectratio=true,width=0.9\textwidth]{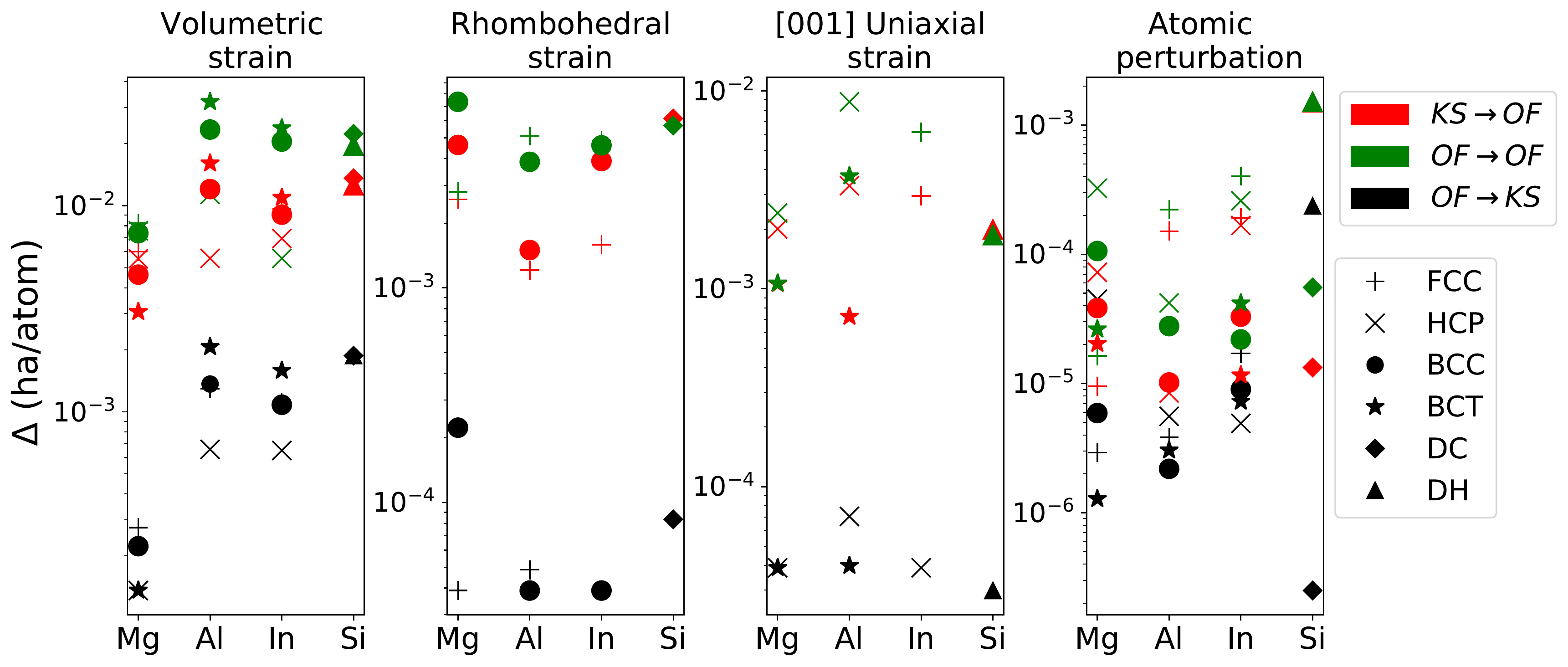}
\caption{\label{Fig:results} $\Delta$-error in the $E_{OF \shortrightarrow KS}$, $E_{KS \shortrightarrow OF}$, and $E_{OF \shortrightarrow OF}$ energies.}
\end{figure*}

In Fig.~\ref{Fig:results} and Table~\ref{table:avgDelta}, we summarize the results so obtained, with the  detailed data available in the Supplementary Material. We observe the following trend in the $\Delta$-errors: $\Delta_{OF \shortrightarrow KS} < \Delta_{KS \shortrightarrow OF} < \Delta_{OF \shortrightarrow OF}$.  In particular, the values of $\Delta_{OF \shortrightarrow KS}$ are quite small, which suggests that the  ground state density in OF-DFT is close to that in KS-DFT. Furthermore, since the values of $\Delta_{KS \shortrightarrow OF}$ are relatively large, it can be inferred that the energy errors in OF-DFT are not a consequence of the errors in the ground state density, but are rather due to a  fundamental limitation in the energy functional.  Indeed, similar trends and inferences follow when comparing physical observables such as equilibrium volume and bulk modulus (Supplementary Material).

Though the current findings cannot be considered to be universal,   we expect them to be generally true for TFW OF-DFT, given that the chosen systems contain several families of materials (metal, semiconductors), classes of structures (close-packed, open and intermediate), symmetries (cubic, hexagonal, tetragonal, orthorhomdic, rhombohedral), and types of distortions (cell, atom). To further bolster our findings, we look to theory, as described below.

%To verify the above hypotheses, we can take recourse to linear response theory: 
The above findings can be understood in terms of linear response theory \cite{martin2020electronic, baroni2001phonons}: 
\begin{widetext}
\begin{align}
\Delta &  E_{KS \shortrightarrow KS}(g)  = E_{KS}(\rho_{KS}(g)) - E_{KS}(\rho_{KS}(0)) \nonumber \\
 & = \int [\rho_{KS}(0)](\mathbf{r}) \Delta V_{ext}(\mathbf{r}) \, \mathrm{d}\mathbf{r} 
+\frac{1}{2} \int \Delta V_{ext}(\mathbf{r}) \tilde{\chi}_{KS}^{-1}(\mathbf{r},\mathbf{r}') \Delta \rho_{KS}(\mathbf{r}') \, \mathrm{d}\mathbf{r} \mathrm{d}\mathbf{r}' 
+\frac{1}{2} \int \Delta \rho_{KS}(\mathbf{r}) \chi_{KS}^{-1}(\mathbf{r},\mathbf{r}') \Delta \rho_{KS}(\mathbf{r}') \, \mathrm{d}\mathbf{r} \mathrm{d}\mathbf{r}' \label{Eq:LR:KSKS} \\
\Delta & E_{OF \shortrightarrow KS}(g)  = E_{KS}(\rho_{OF}(g)) - E_{KS}(\rho_{OF}(0)) \nonumber \\
& = \int [\rho_{OF}(0)](\mathbf{r}) \Delta V_{ext}(\mathbf{r}) \, \mathrm{d}\mathbf{r} 
+\frac{1}{2} \int \Delta V_{ext}(\mathbf{r}) \tilde{\chi}_{KS}^{-1}(\mathbf{r},\mathbf{r}') \Delta \rho_{OF}(\mathbf{r}') \, \mathrm{d}\mathbf{r} \mathrm{d}\mathbf{r}' 
+ \frac{1}{2} \int \Delta \rho_{OF}(\mathbf{r}) \chi_{KS}^{-1}(\mathbf{r},\mathbf{r}')  \Delta \rho_{OF}(\mathbf{r}') \, \mathrm{d}\mathbf{r} \mathrm{d}\mathbf{r}' \label{Eq:LR:OFKS} \\
 \Delta & E_{KS \shortrightarrow OF}(g)  = E_{OF}(\rho_{KS}(g)) - E_{OF}(\rho_{KS}(0)) \nonumber \\
& = \int [\rho_{KS}(0)](\mathbf{r}) \Delta V_{ext}(\mathbf{r}) \, \mathrm{d}\mathbf{r} 
+ \frac{1}{2} \int \Delta V_{ext}(\mathbf{r}) \tilde{\chi}_{OF}^{-1}(\mathbf{r},\mathbf{r}') \Delta \rho_{KS}(\mathbf{r}') \, \mathrm{d}\mathbf{r} \mathrm{d}\mathbf{r}' 
 + \frac{1}{2} \int \Delta \rho_{KS}(\mathbf{r}) \chi_{OF}^{-1}(\mathbf{r},\mathbf{r}')  \Delta \rho_{KS}(\mathbf{r}') \, \mathrm{d}\mathbf{r} \mathrm{d}\mathbf{r}' \label{Eq:LR:KSOF} \\
 \Delta & E_{OF \shortrightarrow OF}(g)   = E_{OF}(\rho_{OF}(g)) - E_{OF}(\rho_{OF}(0)) \nonumber \\
& = \int [\rho_{OF}(0)](\mathbf{r}) \Delta V_{ext}(\mathbf{r}) \, \mathrm{d}\mathbf{r} 
+ \frac{1}{2} \int \Delta V_{ext}(\mathbf{r}) \tilde{\chi}_{OF}^{-1}(\mathbf{r},\mathbf{r}') \Delta \rho_{OF}(\mathbf{r}') \, \mathrm{d}\mathbf{r} \mathrm{d}\mathbf{r}' 
 + \frac{1}{2} \int \Delta \rho_{OF}(\mathbf{r}) \chi_{OF}^{-1}(\mathbf{r},\mathbf{r}')  \Delta \rho_{OF}(\mathbf{r}') \, \mathrm{d} \mathbf{r} \mathrm{d}\mathbf{r}' \label{Eq:LR:OFOF}
\end{align} 
\end{widetext}
\noindent where the perturbation in electron density:
\begin{align}
\Delta \rho_{KS} & = \rho_{KS}(g) - \rho_{KS}(0) \,, \\
\Delta \rho_{OF} & = \rho_{OF}(g) - \rho_{OF}(0) \,,
\end{align}
the perturbation in the external potential
\begin{align}
\Delta V_{ext} = V_{ext}(g) - V_{ext}(0) \,,
\end{align}
and the linear response kernels:
\begin{align}
\chi_{KS}^{-1} = \frac{\delta^2 E_{KS}}{\delta \rho(\mathbf{r}) \delta \rho(\mathbf{r'})} \,, \quad \tilde{\chi}_{KS}^{-1} = \frac{\delta^2 E_{KS}}{\delta V_{ext}(\mathbf{r}) \delta \rho(\mathbf{r'})} \,, \\
\chi_{OF}^{-1} = \frac{\delta^2 E_{OF}}{\delta \rho(\mathbf{r}) \delta \rho(\mathbf{r'})} \,, \quad \tilde{\chi}_{OF}^{-1} = \frac{\delta^2 E_{OF}}{\delta V_{ext}(\mathbf{r}) \delta \rho(\mathbf{r'})} \,.
\end{align} 
The above equations are applicable to any perturbation, with the energy functional assumed to be dependent on the electron density alone, consistent with the Hohenberg-Kohn theorem \cite{PhysRev.136.B864}. Indeed, in the case of strains and atomic perturbations, the changes in energy are dictated by the elastic constants and interatomic force constant matrix, respectively, which can be derived in the context of the above formalism, as done previously in KS-DFT \cite{baroni2001phonons}.

Since the difference between $\Delta E_{KS \shortrightarrow KS}$ and $\Delta E_{OF \shortrightarrow KS}$ is small, as found in the numerical results above, it follows from Eqs.~\ref{Eq:LR:KSKS} and \ref{Eq:LR:OFKS} that the ground state densities of  KS-DFT and OF-DFT are \emph{close}, which also justifies the use of linear response theory. Furthermore, since the difference between $\Delta E_{KS \shortrightarrow OF}$ and $\Delta E_{KS \shortrightarrow KS}$ is relatively large, as found in the numerical results above, it follows from Eqs.~\ref{Eq:LR:KSKS} and \ref{Eq:LR:KSOF} that the error in the energy for OF-DFT calculations is due to the poor representation of the linear response susceptibility $\chi_{OF}$ relative to $\chi_{KS}$, which confirms the need for developing alternate functionals with better linear response.  Indeed, as discussed in Section~\ref{Sec:Introduction}, this has motivated the development of a number of local \cite{luo2018simple, constantin2018semilocal, perdew2007laplacian, constantin2017modified, francisco2021analysis} and nonlocal \cite{chacon1985nonlocal, Igor, wang1992kinetic, wang1999orbital, zhou2005improving, huang2010nonlocal, doi:10.1063/1.5023926, shin2014enhanced, shao2021revised} functionals. Given that these functionals are generally more accurate than TFW \cite{constantin2019performance}, it is expected that the current findings are also applicable to them, though the accuracy of the ground state density and computed energy may vary between different functionals. Since the focus of the current work is to  investigate the source of error in the TFW functional, rather than provide a  comparison between different functionals, we refrain from such comparison here.

\begin{table}[t]
\begin{tabular}{ccccccc}
\hline\hline
&~~& $\Delta_{OF\shortrightarrow KS}$&~~& $\Delta_{KS\shortrightarrow OF}$ &~~& $\Delta_{OF\shortrightarrow OF}$ \\
\hline
Volumetric strain &~~& 0.0010243  &~~& 0.0091202 &~~& 0.0166368 \\
Rhombohedral  Strain &~~&0.0000466 &~~& 0.0028420 &~~&  0.0044053 \\
$[001]$ Uniaxial strain &~~& 0.0000423  &~~& 0.0020935 &~~& 0.0036976 \\  
Atomic perturbation &~~&0.0000246 &~~& 0.0001598 &~~&   0.0002204 \\
\hline\hline
\end{tabular}
\caption{\label{table:avgDelta} Average $\Delta$-error in the $E_{OF \shortrightarrow KS}$, $E_{KS \shortrightarrow OF}$, and $E_{OF \shortrightarrow OF}$ energies.}
\end{table}

The above results also suggest a possible strategy to accelerate KS-DFT calculations without significant loss of accuracy. In particular, the ground state density computed from OF-DFT, which scales linearly with system size, can be used as input to accelerate SCF convergence or just perform a single SCF iteration in KS-DFT. Indeed, to increase the fidelity of such single-shot calculations, nonlocal pseudopotentials, which are the standard in KS-DFT, need to be incorporated into OF-DFT \cite{xu2022nonlocal}.

%%%%%%%%%%%%%%%%%%%%%%%%%%%%%%%%%%%%%%%%%%%%%%
\section{Concluding remarks}
In this work, we have systematically investigated the source of error arising in the TFW density functional relative to Kohn-Sham DFT. In particular, through numerical studies on a variety of materials, for a range of crystal structures subject to strains and atomic displacements,  we have found  that while the ground state electron density in the TFW variant of orbital-free DFT is close to the Kohn-Sham ground state density, the corresponding energy differs significantly from the Kohn-Sham value. We have  shown that these differences arise due to the poor representation of the linear response  within the TFW approximation for the electronic kinetic energy,  therefore confirming conjectures in the literature. In so doing, we have found that the energy computed from a non-self-consistent Kohn-Sham calculation using the TFW ground state density as input is in very good agreement with the energy obtained from the fully self-consistent Kohn-Sham solution. 

The development of more general and accurate  electronic kinetic energy functionals for use in orbital-free DFT,  possibly aided by state-of-the-art machine learning techniques, is therefore a worthy subject of pursuit. 

%%%%%%%%%%%%%%%%%%%%%%%%%%%%%%%%%%%%%%%%%%%

\section{Supplementary Material}
\vspace{-2mm} Plots of variation in $\Delta E_{KS \shortrightarrow KS}$, $\Delta E_{OF \shortrightarrow KS}$, $\Delta E_{KS \shortrightarrow OF}$, and $\Delta E_{OF \shortrightarrow OF}$ with perturbation. 

%%%%%%%%%%%%%%%%%%%%%%%%%%%%%%%%%%%%%%%%%%% 
\section{Acknowledgements}
\vspace{-2mm}The authors would like to thank Sam Trickey for helpful discussions on orbital-free DFT literature. The authors also sincerely acknowledge the invaluable help from Maria Emelianenko. The authors also thank the anonymous reviewers for their valuable comments and suggestions. B.T. acknowledges the support of the Quantum Science and Engineering Center (QSEC) at George Mason University. X.J., P.S., and J.P. acknowledge the support of Grant No. DE-SC0019410, funded by the U.S. Department of Energy, Office of Science. This work was performed, in part, under the auspices of the U.S. Department of Energy by LawrenceLivermore National Laboratory, under Contract No. DE-AC52-07NA27344.

%%%%%%%%%%%%%%%%%%%%%%%%%%%%%%%%%%%%%%%%%%%%

\section*{Data Availability}
\vspace{-2mm} The data that supports the findings of this study are available within the article and its supplementary material.

\section*{Author Declarations} 
\vspace{-2mm} The authors have no conflicts to disclose.

%%%%%%%%%%%%%%%%%%%%%%%%%%%%%%%%%%%%%%%%%%%% bibiliography
% \bibliographystyle{IEEEtran}
\bibliography{bibilio.bib}

\end{document}